\documentclass[12pt]{article}

\usepackage{epsfig}

\usepackage{amsmath}
\usepackage{amssymb}
\usepackage{cite}

\usepackage{a4}

\usepackage{fancyhdr}
\numberwithin{equation}{section}
\numberwithin{figure}{section}
\newcommand{\beqs}{\begin{equation*}}
\newcommand{\beq}{\begin{equation}}

\newcommand{\eeqs}{\end{equation*}}
\newcommand{\eeq}{\end{equation}}

\newcommand{\beqas}{\begin{eqnarray*}}
\newcommand{\beqa}{\begin{eqnarray}}

\newcommand{\eeqas}{\end{eqnarray*}}
\newcommand{\eeqa}{\end{eqnarray}}

\newcommand{\eq}[2]{\begin{equation} #1 \label{#2} \end{equation}}

\newcommand{\eqa}[2]{\begin{eqnarray} #1 \label{#2} \end{eqnarray}}

\newcommand{\meq}[2]{\begin{multline} #1 \label{#2} \end{multline}}

\newcommand{\eps}{\varepsilon}
\newcommand{\al}{\alpha}
\newcommand{\be}{\beta}
\newcommand{\ga}{\gamma}
\newcommand{\de}{\delta}
\newcommand{\om}{\omega}
\newcommand{\ka}{\kappa}
\newcommand{\la}{\lambda}
\newcommand{\si}{\sigma}

\newcommand{\Om}{\Omega}

\newcommand{\blist}{\begin{itemize}}

\newcommand{\elist}{\end{itemize}}

\providecommand{\href}[2]{#2}

\newcommand{\unity}{1\hspace{-0.243em}\text{l}}

\newcommand{\Epo}{E^+_1}
\newcommand{\Emo}{E^-_1}

\begin{document}

%%% TITLEPAGE STARTS HERE %%%

\begin{titlepage}

\renewcommand{\thefootnote}{\fnsymbol{footnote}}

\hfill TUW--03--14

% \hfill Vers. 1.0 -- DRAFT \today\\ %% just for draft versions 

\begin{center}
\vspace{0.5cm}

{\Large\bf Three functions in dilaton gravity: The good, the bad and the muggy}
\vspace{1.0cm}
% \vfill

{\bf D.\ Grumiller\footnotemark[1]
%, W.\ Kummer\footnotemark[2] and D.V.\ Vassilevich\footnotemark[3]\footnotemark[4]
}
\vspace{7ex}

  {\footnotemark[1]%\footnotemark[2]
    \footnotesize Institut f\"ur
    Theoretische Physik, Technische Universit\"at Wien \\ Wiedner
    Hauptstr.  8--10, A-1040 Wien, Austria}
\vspace{2ex}

  %{\footnotemark[3]\footnotesize Intitut f\"{u}r Theoretische Physik, Universit\"{a}t Leipzig,\\Augustusplatz 10, D-04109 Leipzig, Germany}

   \footnotetext[1]{e-mail: \texttt{grumil@hep.itp.tuwien.ac.at}}
   %\footnotetext[2]{e-mail: \texttt{wkummer@tph.tuwien.ac.at}}
   %\footnotetext[3]{e-mail: \texttt{vassil@itp.uni-leipzig.de}}
   %\footnotetext[4]{On leave from V.Fock Institute of Physics,St.Petersburg University, 198904 St.Petersburg, Russia}
\end{center}
\vspace{7ex}

\begin{abstract}

%%% ABSTRACT STARTS HERE %%%

Dilaton gravity in two dimensions is briefly reviewed from the perspective of three dilaton potentials: One determines classical physics (``the good'', denoted by $w$), the second is relevant for semi-classical (and quantum) effects (``the muggy'', denoted by $I$) and the third could be responsible for nonperturbative quantum effects (``the bad'', denoted by $Z$).

This paper is based upon lectures given in Cernowitz in Oc\-to\-ber/No\-vem\-ber 2002 at The XIV International Hutsulian Workshop ``Mathematical Theories and their Physical \& Technical Applications''.
  
%%% END OF ABSTRACT %%%

\end{abstract}

%PACS numbers: 
% 04.60.Kz; 04.60.Gw; 11.10.Lm; 11.80.Et

\vfill
\end{titlepage}

%%% END OF TITLEPAGE %%%

%%% PAPER STARTS HERE %%%

\section{Structure of this paper}

Instead of a proper introduction I will refer to a recent review on dilaton gravity in two dimensions for supplementary reading, motivations why dilaton gravity in two dimensions could be interesting and a summary of some of the results obtained during the last decade \cite{Grumiller:2002nm}. I believe there is no point to copy and paste everything from there to this proceedings contribution. Whenever some details are missing in this work it is suggested to consult the review article. I tried to keep most of the conventions, so there is no mayor incompatibility between that work and the present one.

Still, this paper should be self-contained. Therefore, I chose to present dilaton gravity from a slightly different perspective than in previous publications and to provide some simple examples. Instead of the commonly used potentials present in the action certain (integrated) combinations of them are taken as the basis of the discussion. This is advantageous, because one of the functions (namely ``the good'' $w$) entails most features that are relevant classically and the other one (``the muggy'' $I$) plays an important role semi-classically and at the quantum level. The third function (``the bad'' $Z$) is usually neglected, but it could be relevant for nonperturbative quantum effects.\footnote{Probably I should explain the chosen adjectives: ``bad'' refers to the undesirable fact that for noninvertible $Z(X)$ a straightforward application of first order gravity is not possible; ``good'' has been chosen because $w(X)$ essentially defines the causal structure of the spacetime and generates the vertices for classical scattering; finally, the somewhat unexpected ``muggy'' has three explanations: $I(X)$ appears in the Hawking flux (related to evaporation of something hot), it causes confusion sometimes (related to damp) and, most importantly, to a first approximation it rhymes with the expected ``ugly''.}

A similar discussion could be relevant for higher dimensional scalar tensor theories, e.g.\ fourdimensional ones which have attracted some attention in the early 1960ies due to work by Fierz, Jordan, Brans and Dicke \cite{Fierz:1956} and in the 1990ies (sometimes under the name ``quintessence'') \cite{Wetterich:1988fm} due to evidence for a nonvanishing ``dark energy'' in connection with the observation of supernovae at high values of the redshift \cite{Riess:2001gk}. E.g.\ it would be gratifying to obtain an analogue of the Minkowski ground state condition (eq.\ (\ref{pr:19}) below) or the (A)dS ground state condition.

The paper is organized as follows: in section \ref{se:1} a brief introduction cannot be avoided -- after all the notation has to be fixed and the models under consideration have to be introduced. Section \ref{se:2} is devoted to ``the good'', a dilaton dependent function denoted by $w(X)$, which determines most of the classical properties. In section \ref{se:3} ``the muggy'', denoted by $I(X)$, is discussed -- it determines much of the semi-classical and quantum behavior, but also the way in which test-particles are affected by geometry. Finally, section \ref{se:4} discusses ``the bad'', denoted by $Z(X)$; it is argued to play a role for nonperturbative quantum effects. Appendix \ref{app:2} recalls some results relevant for Hawking radiation. In appendix \ref{app:3} the path integral quantization of generic dilaton gravities with scalar matter is redone; some new one loop results are presented. Appendix \ref{app:4} contains a list of questions that arose during and after the lectures (together with some answers). All appendices contain (hopefully clarifying) remarks which are somewhat scattered in the original papers.

\section{Dilaton gravity in two dimensions}\label{se:1}

Starting point is the action describing generalized dilaton gravity\footnote{More general dilaton gravity actions than (\ref{eq:GDT}) do exist involving higher powers of $(\nabla X)^2$ (cf.\ e.g.\ \cite{Grumiller:2002md} for a particular class of examples relevant for the exact string BH of Dijkgraaf, Verlinde and Verlinde \cite{Dijkgraaf:1992ba}), but most models in the literature are of the much simpler form (\ref{eq:GDT2}). The most popular is the CGHS model \cite{Callan:1992rs,Russo:1992ht,Mandal:1991tz}. Earlier well-known models include the one of Jackiw-Teitelboim \cite{Barbashov:1979bm}, the Katanaev-Volovich model \cite{Katanaev:1986wk} and spherically reduced gravity \cite{Berger:1972pg,Kuchar:1994zk,Grumiller:1999rz}.} (GDT)
\begin{equation}
\label{eq:GDT}
L^{(\textrm{dil})}=\int d^{2}x\, \sqrt{-g}\; \left[ Z(X) \frac{R}{2}-\frac{U(X)}{2}\; (\nabla X)^{2}+V(X)\; \right] +L^{(m)}\, ,
\end{equation}
 where \( R \) is the Ricci-scalar, \( X \) the dilaton, \( Z(X), U(X) \) and \( V(X) \)
arbitrary functions thereof, \( g \) is the determinant of the metric \( g_{\mu \nu } \),
and \( L^{(m)} \) contains eventual matter fields. For the moment we will set $L^{(m)}=0$ and focus solely on the geometric sector. 

If ``the bad'' function $Z(X)$ is invertible\footnote{Some literature based upon \cite{Russo:1992yg} uses $Z\neq X$ and $U=1$. There it is argued that a nontrivial function $U$ can be absorbed into a redefinition of the dilaton, but of course also in that case obstructions may occur.} in the range of definition of $X$ then a new dilaton $\tilde{X}=Z(X)$ can be introduced and one obtains instead \cite{Russo:1992yg,Odintsov:1991qu}
\begin{equation}
\label{eq:GDT2}
\tilde{L}^{(\textrm{dil})}=\int d^{2}x\, \sqrt{-g}\; \left[ \tilde{X} \frac{R}{2}-\frac{\tilde{U}(\tilde{X})}{2}\; (\nabla \tilde{X})^{2}+\tilde{V}(\tilde{X})\; \right] + \tilde{L}^{(m)}\, ,
\end{equation}
with some new potentials $\tilde{U}$ and $\tilde{V}$. To simplify the notation we drop the tilde from now on. The action (\ref{eq:GDT2}) is the starting point of all calculations which involve the first order formulation of gravity 
\begin{equation}
\label{2.62}
L^{\rm (FOG)}=\int \left[ X_{a}(D\wedge e)^{a}+Xd\wedge\omega +\epsilon \left(U(X)X^aX_a+V(X)\right)\right] \; ,
\end{equation}
which seems to have been introduced first for the special case (\( U=V=0 \)) in string theory \cite{Verlinde:1991rf}, then considered for a special model in ref.\ \cite{Ikeda:1993aj} and finally generalized to the in \( D=2 \) most general form for a theory of pure gravity in refs.\  \cite{Strobl:1994yk,Schaller:1994es}.
Most symbols have their ``usual'' meaning: $e^a$ is the zweibein one-form,
$\epsilon$ is the volume two-form. The one-form $\omega$ represents the 
spin-connection $\om^a{}_b=\eps^a{}_b\om$ with $\eps_{ab}$ being the totally
antisymmetric Levi-Civit{\'a} symbol. The action (\ref{2.62}) 
depends on two auxiliary fields $X^a$. Its geometric part is a special case of a 
Poisson-$\si$ model \cite{Ikeda:1993aj,Ikeda:1994fh,Strobl:1994yk,Schaller:1994es}
with a three dimensional target space the
coordinates of which are $X,X^a$. With flat metric $\eta_{ab}$ in light-cone coordinates ($\eta_{+-}=1=\eta_{-+}$, $\eta_{++}=0=\eta_{--}$) the first (``torsion'') term of (\ref{2.62}) is given by
\begin{equation}
X_a(D\wedge e)^a = \eta_{ab}X^b(D\wedge e)^a =X^+(d-\omega)\wedge e^- +
X^-(d+\omega)\wedge e^+\,.\label{dvvXDe}
\end{equation}

Until section \ref{se:4} it will be assumed always that $Z(X)$ is, indeed, invertible in the whole range of definition of the dilaton $X$. This allows to employ the first order formulation (\ref{2.62}), which turned out to be very convenient classically \cite{Klosch:1996fi} and crucial at the quantum level \cite{Haider:1994cw,Kummer:1998zs,Grumiller:2000ah,Grumiller:2001ea}.

Having introduced ``the bad'' $Z(X)$ already in the first formula (\ref{eq:GDT}) it is now proper to define ``the good'' $w(X)$ and ``the muggy'' $I(X)$:
\begin{equation}
w(X):=\int^{X}V(y)I(y)dy\,,
\label{vbh:w}
\end{equation}
\begin{equation}
I(X):=\exp{\int^{X}U(y)dy}.
\label{vbh:IQ}
\end{equation}
These somewhat bizarre combinations of the potentials $U$ and $V$ appear nevertheless ``naturally'' and will be discussed in detail in the next two sections. One important feature will be presented immediately: in all GDTs an absolutely conserved quantity exists\footnote{It is a consequence of the fact that a Poisson-$\si$ model with odd-dimensional target space necessarily implies the existence of (at least) one Casimir function \cite{Grabowski:1993}.} \cite{Kummer:1991bg} even in the presence of matter \cite{Mann:1993yv,Grumiller:1999rz}. In the absence of matter it reads
\eq{
\mathcal{C}^{(g)}=I(X)X^+X^-+w(X)\,,\quad d\mathcal{C}^{(g)}=0\,,
}{pr:1}
thus involving ``the good'' and ``the muggy''. On each classical solution $\mathcal{C}^{(g)}$ is constant and related to the total mass of the system.

Inverting eqs.\ (\ref{vbh:w}) and (\ref{vbh:IQ}) establishes
\eq{
U(X)=\frac{I'(X)}{I(X)}\,,\quad V(X)=\frac{w'(X)}{I(X)}\,.
}{pr:17}
Thus, the action (\ref{2.62}) is symmetric under constant rescalings $I\to\la I, w\to\la w$ with a scale parameter $\la$ and under constant shifts $w\to w+c$. Comparison with (\ref{pr:1}) shows that the former operation corresponds to a rescaling of the mass, while the latter corresponds to a shift.

An interesting two parameter class of models (henceforth ``$ab$-family'') has been discussed in \cite{Katanaev:1997ni}. The potentials read
\eq{
w(X)=-\frac{B}{2(b+1)}X^{b+1}\,,\quad I(X)=X^{-a}\,,\quad a\in\mathbb{R}\,, b\in\mathbb{R}\setminus\{-1\}\,.
}{pr:18}
The scale parameter $B$ is practically irrelevant. Only its sign matters. 

Within the $a-b$ family there is a subclass of models fulfilling the Minkowski ground state (MGS) property\footnote{There exist several equivalent versions of (\ref{pr:19}), e.g.\ $V\propto d/dX I^{-2}$ or $U=-d/dX \ln{w}$.}
\eq{
w\cdot I={\rm const.} \quad\Rightarrow\quad a=b+1\,.
}{pr:19}
An MGS model has, as the name indicates, a Minkowskian ground state, i.e.\ there exists a value of the Casimir function (which can be shifted to zero; therefore we will always assume that $\mathcal{C}^{(g)}=0$ is the ground state solution) such that the corresponding classical solution for the line element is Minkowski spacetime. Other relevant subclasses describe a BH immersed in Rindler space ($w\propto X$) or a (anti-)deSitter ground state theory ($V\propto X$). The CGHS model is the only one that fulfills all three requirements simultaneously.

\section[The function {${w}(X)$} and classical physics]{The function $\boldsymbol{w(X)}$ and classical physics}\label{se:2}

\subsection{Pure geometry}

Locally all classical solutions of (\ref{2.62}) can be found rather easily. The most interesting quantity is the line element. Following a standard procedure it can be presented in an Eddington-Finkelstein like gauge:
\eq{
(ds)^2=2I(X)dfdX+2I(X)\left(\mathcal{C}^{(g)}-w(X)\right)(df)^2\,.
}{pr:2}
If $I(X)$ allows for such a redefinition it is convenient to introduce a new dilaton $\tilde{X}$ defined by $d\tilde{X}=I(X)dX$, thus obtaining
\eq{
(ds)^2=2dfd\tilde{X}+K(\tilde{X})(df)^2\,,\quad K(\tilde{X})=2I(X(\tilde{X}))\left(\mathcal{C}^{(g)}-w(X(\tilde{X}))\right)\,,
}{eq:kn}
with the Killing norm $K(\tilde{X})$. The line element (\ref{pr:2}) is conformally related to
\eq{
(d\tilde{s})^2=2dfdX+2\left(\mathcal{C}^{(g)}-w(X)\right)(df)^2\,.
}{pr:3}
The conformal factor which induces the mapping between (\ref{pr:2}) and (\ref{pr:3}) is given by $\Om=\sqrt{I(X)}$. Obviously the conformal factor is singular if $I$ vanishes or diverges. The target space coordinates and Cartan variables transform as
\eq{
e^a=\Om\tilde{e}^a\,,\quad X^a=\Om^{-1}\tilde{X}^a\,,\quad\om=\tilde{\om}+\frac{U}{2}\tilde{X}^ae^b\eta_{ab}\,,
}{pr:666}
and the line element is given by $ds^2=\Om^2d\tilde{s}^2$, as can be seen by comparing (\ref{pr:3}) with (\ref{pr:2}).

Note that the Casimir function as defined in (\ref{pr:1}) and ``the good'' function are both insensitive to the conformal transformation. Thus, these quantities encode the relevant features of the causal structure (which is not changed by regular conformal transformations \cite{waldgeneral}). 

In particular, the zeros of $(\mathcal{C}^{(g)}-w(X)$ correspond to Killing horizons (provided $I$ does not become singular at these points). Thus, by choosing $w(X)$ and the number of zeros it exhibits black holes (BHs) can be modeled with an arbitrary number of extremal and/or non-extremal horizons. As noted below eq.\ (\ref{pr:17}) $w$ contains a bit more information than the potentials in the action, namely an integration constant: arbitrary constant shifts $w\to w+c$ can be compensated for each solution by $\mathcal{C}^{(g)}\to\mathcal{C}^{(g)}+c$. Therefore, a shift in $w$ typically just corresponds to a shift of the BH mass.

In summary, $w(X)$ encodes the essential features of the causal structure, namely the existence, number and types of Killing horizons. The action (\ref{2.62}) is invariant under constant shifts of $w(X)$.

\subsection{Scalar matter}

Adding, for instance, scalar matter\footnote{Matter contributions of this type appear also in the dimensional reduction of higherdimensional scalar tensor theories. In that case also the scalar field can be interpreted as a ``dilaton'' field, at least from the higherdimensional point of view \cite{Grumiller:2000hp}.}
\eq{
L^{(m)}=\frac{1}{2}\int F(X)d\phi\wedge\ast d\phi\,,
}{pr:6}
has drastic consequences because it adds physical propagating degrees of freedom which allow scattering processes (without matter no such processes do exist). Moreover, it destroys integrability, so at a certain point one is forced to introduce approximative methods like perturbation theory. If the coupling function $F(X)$ is constant we will call the scalar field $\phi$ minimally coupled, otherwise non-minimally.

Applying the path integration procedure presented in appendix \ref{app:3} yields a nonlocal nonpolynomial action depending solely on the matter fields and external sources. The crucial observation is that the term in (\ref{Wtilde-b}) which generates the classical vertices\footnote{Incidentally also the Hawking temperature as calculated from surface gravity is proportional to $w'$ evaluated at the horizon.} is proportional to $w'$. Thus, models with $w(X)=aX+b$ do not produce classical scattering for minimally coupled matter \cite{Grumiller:2002dm}.

What about the asymptotic modes of the scalar field? Inspection of the Klein-Gordon equation (\ref{vbh:matter}) immediately reveals only dependence on $w$ (and for nonminimally coupled matter also on $F$).

Thus, for classical scattering processes only ``the good'' function $w$ is relevant.

\section[The function $I(X)$ and semi-classical physics]{The function $\boldsymbol{I(X)}$ and semi-classical physics}\label{se:3}

The insensitivity of the Killing horizons to the function $I(X)$ together with the conformal invariance of $w(X)$ and with the fact that all metrics in $2D$ are {\em locally} conformally flat has led some authors to the notion of ``conformal invariance of GDTs''. Let us therefore start with a list of quantities which are, indeed, conformally invariant and a second list with non-invariant ones:
\blist
\item {\bf Conformally invariant quantities:} ``the good'' potential $w(X)$, the Casimir function, number and type of Killing horizons, surface gravity, Hawking temperature as derived naively from surface gravity, asymptotic modes, classical vertices, parameter $b$ in the $ab$-family
\item {\bf Conformally non-invariant quantities:} ``the muggy'' potential $I(X)$, curvature as derived from the line element, curvature as defined by the connection, torsion, geodesics, geodesic (in)completeness, volume element, MGS property, Hawking flux, semiclassical vertices, Lagrangian, parameter $a$ in the $ab$-family 
\elist

\subsection{Pure geometry}

Let us focus on the curvature scalar first. There is a slight subtlety at this point, as there are two possible definitions of the curvature scalar (one with torsion, which is the one we are using here and one without torsion): in conformal frames where $I\neq 1$ the curvature scalar (as defined by the Hodge dual of the curvature two form) is not just the second derivative of the Killing norm, because part of the geometric information is encoded in the torsion. In the absence of matter one obtains for the curvature as derived from the Killing norm (\ref{eq:kn})
\eq{
R=2\frac{d^2\xi(\tilde{X})}{d\tilde{X}^2}=2 \left(V'(X)+I^{-1}(X)(\mathcal{C}^{(g)}-w(X))U'(X)\right)-4\frac{w''(X)}{I(X)}\,,
}{err1}
while curvature as defined as the Hodge dual of the curvature 2-form reads\footnote{Curiously if one imposes an analog of the MGS condition on $R^\ast$ rather than on $R$ (i.e.\ one requires that $R^\ast(\mathcal{C}^{(g)}=0)=0$) one obtains the (A)dS ground state condition $V\propto X$.}
\eq{
R^\ast=2 \ast d\wedge\om=-2\left(V'(p_1)+I^{-1}(X)(\mathcal{C}^{(g)}-w(X))U'(p_1)\right)\,.
}{err2}
In the absence of torsion ($U=0$) both definitions coincide as it should be. The torsion 2-form $T^a$ is proportional to the volume 2-form times $U(p_1)t^a$ with $t^+=I(p_1)$ and $t^-=\mathcal{C}^{(g)}-w(p_1)$. From now on we will always mean (\ref{err1}) when talking about ``curvature''. It vanishes for the ground state in MGS models.

Clearly, all quantities which depend in a nontrivial way on curvature will inherit the sensitivity to $I$. The most relevant ones are listed above among the conformally non-invariant quantities.

\subsection{Scalar matter}

Probably the easiest way to see the different roles of $I$ and $w$ is to look at (\ref{Wtilde-b}). All classical vertices are generated by $w'$ (and by $F'$, if the field couples nonminimally). However, one-loop effects are sensitive to $I$, because the auxiliary field $f$ which appears in the measure inherits the $I$ dependence by virtue of (\ref{cghs:4}).

This dependence is expressed, for instance, in the $I$ dependence of the Hawking flux (\ref{1loophaw4}) or in the $I$ dependence of the Polyakov action (\ref{Lpol-gauge}). Physically, the reason is simple: the presence of curvature (which explicitly depends on $I$, see (\ref{err1})) induces a scale and hence generates the conformal anomaly by one loop effects. 

The simplest demonstration of an $I$ dependent observable at one loop level invokes the CGHS model \cite{Grumiller:2003mc}. Depending on the conformal frame one obtains either a vanishing inverse specific heat of the CGHS BH (for $I=1$) or a positive one (for $I=1/X$).

\section[The function $Z(X)$ and quantum physics]{The function $\boldsymbol{Z(X)}$ and quantum physics}\label{se:4}

As mentioned below (\ref{eq:GDT}) if $Z(X):=\tilde{X}$ is globally invertible one can start from the simpler action (\ref{eq:GDT2}) or its first order version (\ref{2.62}). But what if $Z(X)$ is not invertible? There are different possible causes for such a failure: for instance, $Z$ can have singularities prohibiting an inversion. We would like to focus on a very special case of non-invertibility: we assume that $Z$ has one or more extrema. 

The classical equation of motion that is obtained by varying (\ref{eq:GDT}) with respect to the dilaton $X$ reads
\eq{
Z'\frac{R}{2}+V'+U'\frac{(\nabla X)^2}{2}+U\square X=0\,.
}{pr:42}
One can see immediately the problem: for $Z'=0$ the curvature $R$ cannot be extracted anymore from this equation, unless the potentials $U,V$ behave in a very special way to compensate for this zero. More concretely, if $V'+U'(\nabla X)^2+U\square X \neq 0$ at the positions where $Z'=0$ then curvature becomes singular at these points.

Between two adjacent extrema well-known methods can be applied to obtain the classical solutions. So the question remains how to patch them together. To this end it is sufficient to perform a local analysis around one extremum. In fact, we can assume that the extremum is a minimum, because if it is a maximum we can redefine $Z\to-Z$, $U\to-U$ and $V\to -V$ without changing the classical equations of motion. Moreover, we can shift the minimum to $Z=0$ because ${\rm const}\cdot R$ generates only a surface term. Finally, by a constant shift of the dilaton the minimum can be assumed to be located at $X=0$ (such a shift of course changes the $X$ dependence of the other potentials). Thus, a rather canonical example is given by $Z(X)=X^2$. Therefore, let us study models of this type in some detail.

There are two patches, $X>0$ and $X<0$. In the former we can redefine $\sqrt{X_>}:=X$, in the latter $\sqrt{X_<}:=-X$. From now on the indices $>$ ($<$) correspond to patches $X>0$ ($X<0$), respectively. Thus, we obtain for the action
\eq{
L_>=\int d^2x\sqrt{-g}\left[X_>R-U(\sqrt{X_>})(\nabla\sqrt{X_>})^2+2V(\sqrt{X_>})\right]
}{pr:22}
and
\eq{
L_<=\int d^2x\sqrt{-g}\left[X_<R-U(-\sqrt{X_<})(\nabla\sqrt{X_<})^2+2V(-\sqrt{X_<})\right]\,.
}{pr:23}
The corresponding solutions for the line element,
\eq{
(ds)^2_{>,<}=2I(\pm\sqrt{X_{>,<}})dX_{>,<}du+2I(\pm\sqrt{X_{>,<}})\left(\mathcal{C}_{>,<}-w(\pm\sqrt{X_{>,<}})\right)(du)^2\,,
}{pr:24}
have to be patched together at $X_>=0=X_<$. Note that in principle the conserved quantity can be chosen independently in both patches. If the line element is regular at the origin then continuity imposes $\mathcal{C}_{>}=\mathcal{C}_{<}$. In the singular case other matching prescriptions are possible.

It is useful to introduce effective quantities
\eq{
U^{\rm eff}:= - U(\pm\sqrt{X_{>,<}})\frac{1}{4X_{>,<}}\,,\quad V^{\rm eff}=V(\pm\sqrt{X_{>,<}})
}{pr:25}
from which the effective potentials $I^{\rm eff}$ and $w^{\rm eff}$ can be deduced. If $U$ is an even function then $I_>^{\rm eff}=I_<^{\rm eff}$. If $U$ is odd then $I_>^{\rm eff}=1/I_<^{\rm eff}$. For even $U$ and even $V$ we observe $w_>^{\rm eff}=w_<^{\rm eff}$, while for even $U$ but odd $V$ we obtain $w_>^{\rm eff}=-w_<^{\rm eff}$.

A simple and yet illuminating class of examples is given by $Z=X^2$, $U=\tilde{a}$ and $V=\la X^{\tilde{b}}$. The potential $U$ is even, but $V$ can be even or odd depending on $\tilde{b}$. We obtain immediately $I^{\rm eff}\propto X_{>,<}^{\tilde{a}/4}$ and $w^{\rm eff}=\int^{X_{>,<}}z^{\tilde{a}/4}\la(\pm\sqrt{z})^{\tilde{b}}dz$. Thus, with the redefinitions
\eq{
a:=-\frac{\tilde a}{4}\,,\quad b:=\frac{\tilde a + 2\tilde b}{4}\,,\quad B:=-2\la(\pm 1)^{\tilde b}\,,
}{pr:26}
we are almost back at the $ab$-family. Almost, because we have actually one such model in each patch. If $\tilde{b}$ is even then $B_>=B_<$, but for odd $\tilde{b}$ the scale parameter $B$ changes its sign. In the latter case if we choose $\mathcal{C}_>=-\mathcal{C}_<$ then the Killing norm effectively reverses its sign. Such a sign change can be compensated by a sign flip of $I$, which allows for the following interpretation: one has patched to the ``positive radius region'' the ``negative radius solution'' (this is in fact well-known in the context of Schwarzschild geometry: a negative mass solution can be reinterpreted as a positive mass solution with negative radii; the patching corresponds to gluing the naked singularity patch to the Schwarzschild patch -- cf.\ e.g.\ fig.\ 2b in \cite{Hartle:1976tp}).

While classically this patching seems somewhat trivial, in the framework of path integral quantization it becomes essential whether we integrate over all $X$ or over all $X_>$ (or all $X_<$).

That these considerations really might play a role for quantization can be demonstrated by looking at the RST model\footnote{Another more recent instant where a nontrivial $Z(X)$ appeared in the context of quantization of (dimensionally reduced) gravity is ref.\ \cite{Niedermaier:2003fz} (see equation (2.6) of that paper). A generalization of the RST (which is also exactly soluble) with nontrivial $Z(X)$ is given by ref.\ \cite{Zaslavsky:1998ca}.} \cite{Russo:1992ht}. Neglecting the matter part and translating it to our notation it reads
\eq{
Z=X-\frac{\ka}{4}\ln{X}\,,\quad U(X)=-1/X\,,\quad V(X)=-2\la^2X\,.
}{pr:27}
For $\ka=0$ it coincides with the CGHS model (actually, the CGHS model was the basis of the RST model which is obtained from the former by a suitable modification of the semiclassical action). For large values of the dilaton the difference between CGHS and RST is negligible as the linear term in $Z$ dominates. However, at $X=\ka/4$ this ``bad'' function has a minimum and thus our previous discussion applies. The inverse $X=X(\tilde{X})$ is essentially given by the Lambert $W$ function \cite{Corless:1996}, which has two real branches. 

Further insight can be gained by discussing a modified first order gravity action
\eq{
L^{\rm (FOG)}=\int \left[ X_{a}(D\wedge e)^{a}+Y(X)d\wedge\omega +\epsilon {\cal V}(X^aX_a,X)\right] \; ,
}{pr:43}
where the only differences as compared to (\ref{2.62}) are the function $Y(X)$ appearing in front of the curvature term $d\wedge\om$ (it plays the same role as $Z(X)$ in (\ref{eq:GDT})) and the more general dependence on a function ${\cal V}(X^aX_a,X)$ instead of the special case ${\cal V}=U(X)X^+X^-+V(X)$ (this generalization is needed, for instance, in the context of the exact string BH). The modified equations of motion read
\begin{eqnarray}
 &  & Y'(X)dX+X^{-}e^{+}-X^{+}e^{-}=0\, ,\label{eq:a5} \\
 &  & (d\pm \omega )X^{\pm }\mp \mathcal{V}e^{\pm }=0\, ,\label{eq:a6} \\
 &  & Y'(X)d\wedge\omega +\epsilon \frac{\partial \mathcal{V}}{\partial X}=0\, ,\label{eq:a7} \\
 &  & (d\pm \omega )\wedge e^{\pm }+\epsilon \, \frac{\partial \mathcal{V}}{\partial X^{\mp }}=0\, .\label{eq:a8} 
\end{eqnarray}
Thus, only (\ref{eq:a5}) and (\ref{eq:a7}) receive corrections due to $Y(X)$. Again it is obvious that for $Y'=0$ curvature singularities may occur (at least for curvature as defined as the Hodge dual of $d\wedge\om$), unless ${\cal V}$ behaves in a very special way at these points. One can transform to an equivalent second order formulation as outlined in sect.\ 2.2 of \cite{Grumiller:2002nm} and one obtains
\eq{
L^{(\textrm{dil})}=\int d^{2}x\, \sqrt{-g}\; \left[ Y(X) \frac{R}{2}+{\cal V}(-(\nabla Y(X))^2,X)\; \right]\,.
}{pr:44}
Thus, $Y$ can be identified with the bad potential $Z$. For ${\cal V}=X^+X^-U(X)+V(X)$ one finds again a conserved quantity obeying (\ref{pr:1}), however the relation of $I$ and $w$ to $U$ and $V$ is different now:
\eq{
U(X)Y'(X)=\frac{I'(X)}{I(X)}\,,\quad V(X)Y'(X)=\frac{w'(X)}{I(X)}
}{pr:45}
All classical solutions can be found and they are given by (compare with (\ref{pr:2}))
\eq{
(ds)^2=2I(X)\left(Y'(X)dfdX+(\mathcal{C}^{(g)}-w(X))df^2\right)\,.
}{pr:46}
Obviously for $Y'=0$ the determinant of the metric vanishes unless simultaneously $I(X)\to\infty$. But in that case the Killing norm diverges for a generic solution. Rewriting (\ref{pr:46}) as $(ds)^2=2I(X(Y))\cdot(dfdY+(\mathcal{C}^{(g)}-w(X(Y)))df^2)$ shows that $Y$ can be eliminated by allowing for multi-valuedness of $w$ and $I$. For instance, in the previous example $Y=X^2$ and thus one has to introduce $I(\pm\sqrt{X})$ and $w(\pm\sqrt{X})$ as in (\ref{pr:24}). In this sense the ``bad'' potential can always be eliminated, but it is maybe easier to keep it and to restrict oneself to single valued potentials.\footnote{If, on the other hand, $I$ and $w$ are already multi-valued functions which cannot be transformed to single valued ones by introducing a nontrivial $Y(X)$ there seems to be no reason to bother with the ``bad'' potential.}

A possible application could be an attempt to circumvent the no-go result concerning the nonexistence of a dilaton gravity action for the exact string BH \cite{Grumiller:2002md}. The hope of being able to achieve this is based upon the observation that $Y(X)$ provides an additional free function in the action and hence it could be tuned such that both the metric and the dilaton come out correctly. However, applying the same arguments as in appendix A of ref.\ \cite{Grumiller:2002md} shows that dilaton shift invariance restricts to potentials of the type ${\cal V}=Y(X)U(X^+X^-/Y^2(X))$. The precise form of $Y(X)$ is completely irrelevant. Thus, one can simply replace in all formulas of ref.\ \cite{Grumiller:2002md} $X$ by $Y(X)$ and in this manner unfortunately the no-go result is recovered. 

In conclusion, to get more insight into the nonperturbative regime of dilaton gravity (with matter) at the quantum level it seems to be relevant to consider the possibility of a nontrivial ``bad'' function $Z(X)$. It will be interesting to see how the quantization procedure presented in appendix \ref{app:3} is modified, but I will leave this issue for future work.\footnote{If one supposes that nonperturbative effects changing $Z(X)$ have to modify it such that eventual singularities disappear then either $Z$ has to be globally invertible (and thus it becomes irrelevant) or the other potentials have to conspire such that no curvature singularities appear at $Z'=0$.}

\section*{Acknowledgment}

This work has been supported by project P-14650-TPH of the Austrian Science 
Foundation (FWF). I am grateful to W.\ Kummer and D.V.\ Vassilevich for a long time collaboration on $2D$ dilaton gravity and for helpful discussions on this paper. These lectures have been presented at the XIV~$^{\rm th}$ International Hutsulian Workshop on ``Mathematical Theories and their Physical \& Technical Applications'' in Cernivtsi (Ukraine), and I am grateful to the organizers and participants of this meeting for their hospitality and numerous interesting conversations, in particular to L.\ Adamska, S.\ Moskaliuk, S.\ Rumyantsev, A.\ Vlassov and M.\ Wohlgenannt. 

\begin{appendix}

\section{Hawking radiation for MGS models}\label{app:2}

Hawking radiation as defined by surface gravity turns out to be conformally invariant: $T_H\propto \dot{\xi}(X=X_h)$, where $X_h$ is the value of the dilaton at the horizon and the proportionality factor is just a numerical one. Prime denotes derivative with respect to the dilaton $X$ and dot derivative with respect to the rescaled dilaton $\tilde{X}$ where $d\tilde{X}=I(X)dX$. The Killing norm in our notation reads ($M=-\mathcal{C}$)
\eq{
\xi=I(X)\left(-M-w(X)\right)\,.
}{1loophaw1}
The horizon is reached for $w(X)=-M$. From (\ref{vbh:w}),(\ref{vbh:IQ}) one obtains
\eq{
\dot{\xi}=\frac{d\xi}{d\tilde{X}}=U(X)\left(-M-w(X)\right)-w'(X)\,,
}{1loophaw2}
and (cf.\ eq.\ (\ref{err1}))
\eq{
\ddot{\xi}=\left(U'(X)(-M-w(X))-U(X)w'(X)-w''(X)\right)I^{-1}(X)\,.
}{1loophaw3}
The flux component of the energy momentum tensor (in conformal gauge) reads (taken directly from (6.26) of \cite{Grumiller:2002nm})
\eq{
T_{--}=\frac{1}{96\pi}\left[(M+w)^2(2U'-U^2)+2(M+w)w''-(w')^2\right]+t_-\,.
}{1loophaw4}
If the flux is required to vanish for $M=0$ one establishes for the $a-b$ family the MGS condition $a=b+1$. 

The integration constant $t_-$ is fixed by the requirement that $T_{\mu\nu}$ in global coordinates should be finite at the horizon. This translates into $T_{--}(X=X_h)=0$, and thus
\eq{
t_-=\left. \frac{1}{96\pi}(w')^2 \right|_{w(X)=-M}\,.
}{1loophaw5}
Note that the integration constant and the second and third contribution to the Hawking flux are conformally invariant, while the first contribution (the one with the double zero at the horizon) is not. This implies that Hawking temperature derived in this way is indeed conformally invariant since it is proportional to (\ref{1loophaw2}) evaluated at the horizon, i.e.\ to the conformally invariant quantity $w'(X)$.

Near the horizon ($X=X_h+\de X$) the flux component behaves like
\eq{
T_{--}=\frac{(\de X)^2}{96\pi} \left[(w')^2(2U'-U^2)+w'''w'\right]_{X=X_h}+{\cal O}(\de X)^3\,.
}{1loophaw6}
Clearly, the leading order term crucially depends on the chosen conformal frame. The non-invariant part vanishes only in the ``preferred'' frame $U=0$ and in the ``curious'' frame $U=-2/X$. The latter separates different phases of Carter-Penrose-diagrams in the $a-b$ family of theories. The invariant part vanishes for $w=AX^2+BX+C$, including all $a-b$ models with $b=0$ (e.g.\ CGHS) or $b=\pm 1$ (e.g.\ JT). Incidentally, for these models the non-local contribution in (\ref{Lsig2}) also vanishes. For the $a-b$ family eq.\ (\ref{1loophaw6}) reduces to
\eq{
T_{--}=\frac{(\de X)^2}{96\pi} \frac{B^2X^{2(b-1)}}{4} \left[a(2-a)+b(b-1)\right]+{\cal O}(\de X)^3\,.
}{1loophaw6a}

Since the Hawking flux in principle is an observable quantity we have conformal non-invariance of (at least) one observable and hence conformal non-invariance at semi-classical level (despite the invariance of classical S-matrix elements and the invariance of the naively calculated Hawking temperature). 

Now the question arises whether or not there is a ``correct'' conformal frame, in any meaningful sense. One possible criterion is stability of the vacuum, i.e.\ the absence of tadpoles (\ref{stabM}). Applying (\ref{GamEE}) to a given background in a given conformal frame yields
\eq{
\left(\partial_1+(M+w)\partial_0+w'\right)\partial_0\ln{I}=-w''\,.
}{1loophaw7}
With $X=x^0$ and $I'=IU$ we get
\eq{
w''+w'U+U'(w+M)=0\,.
}{1loophaw8}
For a given classical model $w$ and a given ground state (typically one wants to achieve $M(\rm ground state)=0$) one can extract a one-parameter family of potentials $U$ solving (\ref{1loophaw8}). For this ``natural'' ground state the result is
\eq{
U(X)=-\frac{\tilde{c}+w'(X)}{w(X)}\,,\quad \tilde{c}\in\mathbb{R}\,.
}{1loophaw9}
For $c=0$ this is nothing but the MGS condition (\ref{pr:19}) in disguise. Within the $a-b$ family this equation reads explicitly
\eq{
U(X)=\frac{c-(b+1)X^b}{X^{b+1}}\stackrel{!}{=}-\frac{a}{X}\,.
}{1loophaw10}
For $b=0$ and $b=-1$ a one-parameter family of possible potentials is obtained, $U=(c-1)/X$ and $U=c$, respectively. For all other models, only the choice $c=0$ allows for a monomial form of $U(X)$. The stability condition then reads $a=b+1$, i.e.\ we recover the MGS condition.

Thus, the ``correct'' conformal frame generically has to fulfill the MGS condition (\ref{pr:19}), provided the stability condition (\ref{stabM}) is required. The only possible exceptions are ``Rindler ground state models'' with $b=0$ (the Fabbri-Russo family \cite{Fabbri:1996bz}, including the CGHS) or $b=-1$ (classically trivial models with $w=\rm const.$; note however, that in (\ref{pr:18}) this value for $b$ explicitly has been excluded).\footnote{The special role of the models with $b=0,-1$ can be seen directly in the expression for the curvature scalar, eq.\ (3.68) of \cite{Grumiller:2002nm}: for $b=0$ the term which is independent of the Casimir vanishes identically, while for $b=-1$ (after rescaling $B$ with $b+1$ to get rid of the singular denominator) the term does not vanish, but it obeys the same power law in the dilaton as the Casimir dependent one; thus, in the latter case the second term can be absorbed by a constant shift of $\mathcal{C}$.}

\section{Integrating out geometry}\label{app:3}

This appendix contains a very brief summary of the path integral quantization of dilaton gravity with scalar matter and applications to scattering processes. For details it is suggested to consult some of the original literature \cite{Haider:1994cw,Kummer:1998zs,Grumiller:2000ah,Grumiller:2001ea}.
\begin{enumerate}
\item Starting from (\ref{2.62}) and (\ref{pr:6}) a Hamiltonian analysis is performed. One encounters six first class constraints (at each spacepoint), three of which are primary.
\item The constraint algebra closes on $\de$-functions, rather than on $\de'$ as in the Virasoro representation (cf.\ e.g.\ eqs.\ (E.29-E.31) and (E.39) of \cite{Grumiller:2001ea}). Nonminimal coupling $F\neq \rm const.$ induces nontrivial deformations of the structure functions.
\item BRST analysis is performed a la BVF \cite{Fradkin:1975cq}. The BRST charge is nilpotent already at the same level as in Yang-Mills theories despite of the nonconstancy of the structure functions; thus, no higher-order ghost terms arise as compared to Yang-Mills theory (this has to do with the peculiar properties of first order gravity; for similar reasons no ordering ambiguities arise).
\item The gauge fixing fermion is chosen such that Sachs-Bondi gauge is recovered because it has some very convenient properties. All (anti-)ghosts and auxiliary fields are integrated out exactly; they produce some (Faddeev-Popov like) functional determinants. Also the integration over the canonical momentum of the scalar field can be performed exactly as it is just of Gaussian type.
\item The measure of the path integral is fixed such that in the semiclassical limit the proper measure is reproduced \cite{Fujikawa:1988ie}. 
\item After taking care of the measure all remaining geometric fields can be integrated out exactly as they appear only linearly in Sachs-Bondi gauge; the ensuing $\de$-functions can be used to integrate out the target space coordinates $X,X^\pm$. The corresponding functional determinant just cancels the Faddeev-Popov determinant.
\item The result is a generating function for Green functions depending solely on the matter field and external sources:
 \eq{
W[\si]=\left.\int(\mathcal{D}f)\delta\left(f-\frac1i \frac{\delta}{\delta j_3}\right) \widetilde W \right|_{j_3=0}\,,
}{cghs:4}
with
\eq{
\widetilde{W}[f,\si,j_3] = \int (\mathcal{D}\widetilde \phi) \exp i L^{\mbox{\scriptsize{eff}}}
}{Wtilde}
The auxiliary field $f$ has been introduced to take care of the correct path integral measure $(\mathcal{D}\widetilde \phi)=(\mathcal{D}f^{1/2}\phi)$.
\item Some integration constants have to be chosen as homogeneous contributions to the solutions of $X,X^\pm$ in terms of the scalar field. This amounts to fixing the residual gauge freedom (in pre-brane-revolutionary QFT such contributions are usually swept under the rug by imposing natural boundary conditions; however, ``natural'' boundary conditions on the vielbeine would be very unnatural, as the metric would have to vanish; thus, in gravity these boundary terms play an important role). Assuming an asymptotic observer who measures a fixed BH mass $M$ essentially determines all integration constants uniquely. The action appearing in (\ref{Wtilde}) then reads\footnote{An important subtlety should be mentioned: among others, the term $w'(\hat{X})$ arises in a very peculiar manner, namely as a homogeneous contribution coming from an ambiguity in the definitions of sources for the target space coordinates $X,X^\pm$ \cite{Kummer:1998zs,Grumiller:2001ea}. If this ambiguity is not taken into account then no classical vertices arise, but then the expectation values of the vielbeine would not coincide with their classical values (in fact, they would vanish).} 
\meq{
L^{\mbox{\scriptsize{eff}}} =  \int d^2x \Big[F(\hat{X})\left((\partial_0 \phi)(\partial_1 \phi) + (M+w(X))(\partial_0 \phi)^2\right) \\
+ j_3 I(\hat{X}) - w'(\hat{X}) + \phi\si\Big]\,, 
}{Wtilde-b}
with
\eq{
\hat{X}=X+{\cal O}(\phi^2)\,.
}{pr:20}
Only the source $\si$ of the scalar field has been kept together with the source $j_3$ for the vielbein component $e_1^+$ (which in the chosen gauge is nothing but $\sqrt{-g}$). Note that the omitted terms in (\ref{pr:20}) are in general complicated nonlocal expressions. At classical level all nontrivial interactions are generated by $w'(\hat{X})$ (and $F'(\hat{X})$), while quantum corrections are induced by $I(\hat{X})$.
\item Now the $\phi$ integration can be performed in a perturbative manner, i.e.\ one expands $\hat{X}$ (and all functions thereof) in powers of $\phi^2$ (actually in powers of $(\partial_0\phi)^2$); then the quadratic part is isolated and the higher order terms are replaced by functional derivatives with respect to the source $\si$.
\item Finally, the auxiliary field $f$ has to be eliminated. This is straightforward as the functional $\de$-function in (\ref{cghs:4}) just replaces $f$ by $I(\hat{X})$. To leading order (i.e.\ without backreactions upon the measure) the classical result $f=\sqrt{-g^{\rm cl.}}$ is recovered. 
\end{enumerate}
The main difference of our background independent approach as compared to quantization on a fixed background is that geometry is integrated out first (before the matter field(s)). It can be reconstructed unambiguously (including backreactions) order by order in $(\partial_0\phi)^2$.

\subsection{Classical vertices and S-matrix}

Several technical subtleties have not been mentioned so far. The most relevant is how to deal with the nonlocalities that appear inevitably. For the classical vertices a trick can be applied which circumvents these difficulties. In this way the lowest order tree graph vertices split into a symmetric one 
\meq{
V_{a}^{(4)}=-2\int_x\int_y (\partial_0\phi)^2(x)(\partial_0\phi)^2(y)\theta(x^0-y^0)\de(x^1-y^1)F(x^0)
F(y^0)\\
\Bigg[4\left(w(x^0)-w(y^0)\right)-2(x^0-y^0)\Big(w'(x^0)+w'(y^0)\\
+\frac{F'(y^0)}{F(y^0)}w(y^0)+\frac{F'(x^0)}{F(x^0)}\left(w(x^0)+M
\right)\Big)\Bigg]\,.
}{vbh:sym}
and a non-symmetric one
\eq{
V_{b}^{(4)}=-4\int_x\int_y (\partial_0\phi)^2(x)(\partial_1\phi\partial_0\phi)(y)\de(x^1-y^1)F(x^0)F'(y^0)
\left|x^0-y^0\right|\,.
}{vbh:asy}
\begin{figure}
%\epsfxsize=7cm
%\centerline{\epsfbox[70 210 540 360]{bothphi.epsi}}
\epsfig{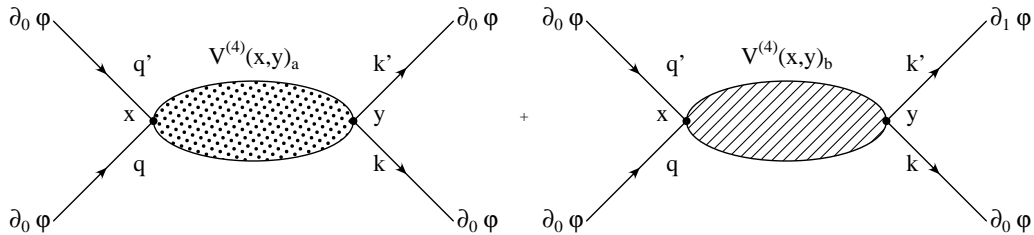}
\caption{The total $V^{(4)}$-vertex (with outer legs) contains a symmetric
contribution $V^{(4)}_a$ and (for non-minimal coupling) a non-symmetric one
$V^{(4)}_b$. The shaded blobs depict the intermediate interactions with VBHs.}
\label{fig1}
\end{figure}

\begin{figure}
\center
\epsfig{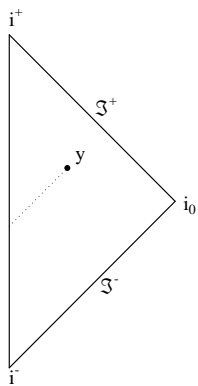}
\caption{Carter-Penrose diagram of VBH geometry}
\label{fig:cp}
\end{figure}
The intermediate states can be interpreted as virtual BHs (VBHs) \cite{Grumiller:2000ah,Fischer:2001vz}. The corresponding Carter-Penrose diagram is depicted in fig.\ \ref{fig:cp}. Note that geometry is trivial apart from the light-like cut. To evaluate the S-matrix all possible geometries of this type are summed coherently (with weight factors following from the theory).

In ref.\ \cite{Grumiller:2002dm} the lowest order vertices for scalar particles scattered on virtual black holes created by their own gravitational self-energy have been calculated (for all GDTs of the form (\ref{2.62})). This is an important first step for obtaining the corresponding $S$-matrix.  

The asymptotic modes of the scalar field obey a Klein-Gordon like equation
\eq{
\partial_0\left(\partial_1 + (M+w(x^0))\partial_0\right) \phi = - \frac{F'(x^0)}{2F(x^0)}\partial_1 \phi - (M+w(x^0)) \frac{F'(x^0)}{F(x^0)}\partial_0 \phi \,.
}{vbh:matter}
Obviously for minimal coupling the right hand side vanishes.

Now some standard techniques\footnote{For details see appendix F of \cite{Grumiller:2001ea}.} can be applied to evaluate the S-matrix $S=\unity+iT$ with
\eq{
T(q, q'; k, k') = \frac{1}{2}\left< 0 \left| a^-_ka^-_{k'} \left(V^{(4)}_a 
+ V^{(4)}_b \right) a^+_qa^+_{q'}\right| 0 \right>\,. 
}{Q129}
The creation and annihilation operators $a^\pm_k$ obey the usual commutation relations $[a_k^-,a_{k'}^+]\propto\de(k-k')$ (with some properly chosen normalization factor) and constitute the building blocks of the Fock space. In and out vacua are identical in the discussed scenario. For gravitational scattering of $s$-waves one obtains in this way a remarkably simple result (which is discussed in more detail in ref.\ \cite{Fischer:2001vz}):
 \eq{
T(q, q'; k, k') = -\frac{i\ka\de\left(k+k'-q-q'\right)}{2(4\pi)^4
|kk'qq'|^{3/2}} E^3 \tilde{T}
}{RESULT}
with the total energy $E=q+q'$,
\eqa{
&& \tilde{T} (q, q'; k, k') := \frac{1}{E^3}{\Bigg [}\Pi \ln{\frac{\Pi^2}{E^6}}
+ \frac{1} {\Pi} \sum_{p \in \left\{k,k',q,q'\right\}}p^2 \ln{\frac{p^2}{E^2}} 
\nonumber \\
&& \quad\quad\quad\quad\quad\quad \cdot {\Bigg (}3 kk'qq'-\frac{1}{2}
\sum_{r\neq p} \sum_{s \neq r,p}\left(r^2s^2\right){\Bigg )} {\Bigg ]},
}{feynman}
and the momentum transfer function\footnote{The square of the momentum 
transfer function is similar to the product of the 3 Mandelstam variables 
$stu$ -- thus we would have non-polynomial terms like $\ln{(stu)^{stu}}$ in 
the amplitude, which is an interesting feature. However, the usual Mandelstam 
variables are not available here, since we do {\em not} have momentum 
conservation in our effective theory (there is just one $\de$-function of
energy conservation).} $\Pi = (k+k')(k-q)(k'-q)$. The 
interesting part of the scattering amplitude is encoded in the scale 
independent (!) factor $\tilde{T}$. 

\begin{figure}
\centering
  \epsfig{file=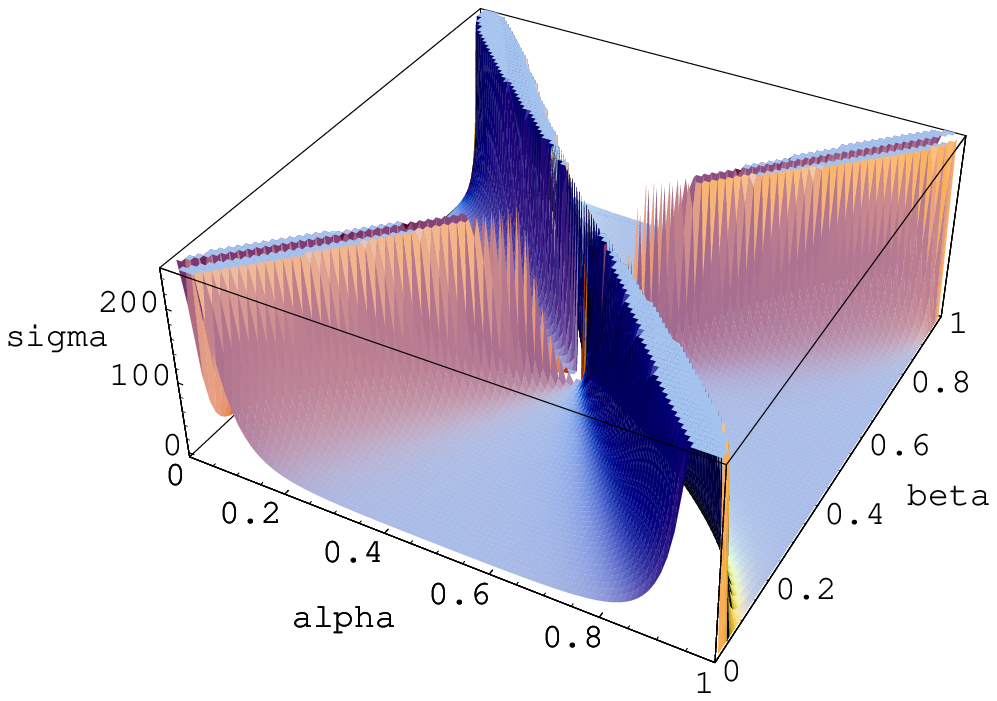,width=.47\linewidth}
  \caption{Kinematic plot of $s$-wave cross-section $d\si/d\al$}
  \label{fig:kin}
\end{figure}

From (\ref{RESULT}) one can, for instance, deduce probabilities of scattering of two incoming $s$-waves with momenta $q,q'$ into two outgoing ones with momenta $k,k'$. Similarly, decay rates of one $s$-wave into three outgoing ones can be calculated.  A typical plot of a cross-section like quantity is depicted in fig.\ \ref{fig:kin}.
On the vertical axis $d\si/d\al$ is plotted, where $\si$ is the total cross section. The two horizontal axes are given by the ingoing ($\al$) and outgoing ($\be$) momenta distributions. If, say, $\al=0.3$ then $30\%$ of the ingoing energy is localized on the ``first'' ingoing $s$-wave and $70\%$ on the ``second''. Since the particles are indistinguishable the plot is symmetric with respect to $\al\to 1-\al$ and $\be \to 1-\be$. One can see clearly the forward scattering poles that occur if the momentum transfer function $\Pi$ has a zero.

Analogous results can be obtained for all other GDTs, but since on the one hand the calculations are somewhat lengthy and on the other hand there is no experimental data for comparison available this has not been done yet.

\subsection{One loop calculations}

\begin{figure}
\begin{center}
\epsfig{file=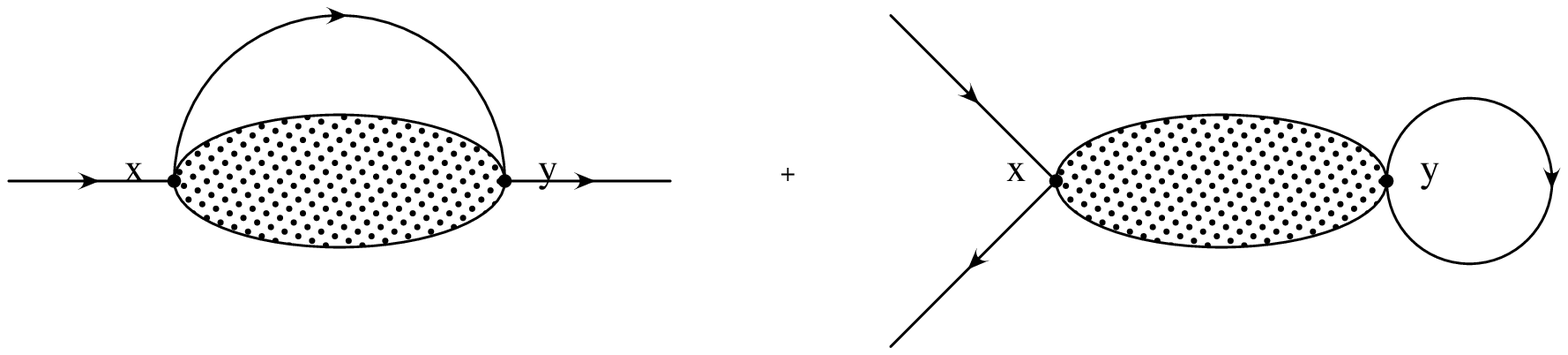, width=\linewidth}
\end{center}
\caption{Self energy}
\label{fig3}
\end{figure}

So far within our approach only the CGHS model has been treated in some detail at one loop level. The result of the one loop corrections to the first nontrivial order\footnote{This corresponds to the graphs shown in fig.\ \ref{fig3}.} in $\phi$ already provides a surprise: the inverse specific heat for the CGHS vanishes, if calculated in the framework of quantization on a fixed background. However, backreactions imply a nontrivial correction, i.e.\ to leading order the specific heat becomes positive and proportional to the square of the mass of the BH \cite{Grumiller:2003mc}.

It would be very interesting to perform the same steps for the Schwarzschild BH. There are two difficulties that one will encounter: first of all, some of the simplifications that occurred in the CGHS case are far from being generic, i.e.\ the calculations will be more involved; secondly (and more important) the coupling to matter will be nonminimal and thus one has to be very careful to obtain the correct one loop effective action. 

For the simpler case of minimal coupling calculations are straightforward.\footnote{The following paragraphs contain unpublished material that has been obtained in collaboration with W.\ Kummer and D.V.\ Vassilevich.} E.g.\ one may use the background field formalism: Let us split the scalar field as
\begin{equation}
\phi=\phi_0+\de\phi \,, \label{split}
\end{equation}
where $\phi_0$ will be regarded as a background field,
$\de\phi$ will represent quantum fluctuation. As usual,
to obtain the one-loop diagrams one should separate
the $\de\phi^2$ order in (\ref{Wtilde-b}):
\meq{
{\cal L}[\de\phi^2]=(\partial_0 \de\phi )(\partial_1 \de\phi )
-(\partial_0 \de\phi )^2 \nabla_0^{-2}
\left[ -w \right]''(\phi_0) \\
+2 [(\partial_0 \de\phi )(\partial_0 \phi_0)]_x
\left( \nabla_0^{-2}
\left[ -w \right]'''(\phi_0)\nabla_0^{-2} \right)_{xy}
[(\partial_0 \de\phi )(\partial_0 \phi)]_y
}{Lsig2}
The first two terms in (\ref{Lsig2}) generate the Polyakov
loop on a background of an effective $\phi_0$-dependent metric.
The second line of (\ref{Lsig2}) contains the terms with
$\de\phi$ taken at different points. Such terms have to be
considered separately; they produce, for instance, the non-local loop shown
in the first graph of fig.\ \ref{fig3}. On a sidenote, the inverse differential operator $\nabla_0^{-2}$ has to be defined properly and the boundary contributions have to be taken into account correctly. They lead, for instance, to the term proportional to $M$ in (\ref{Wtilde-b}).

\begin{figure}
\begin{center}
\epsfig{file=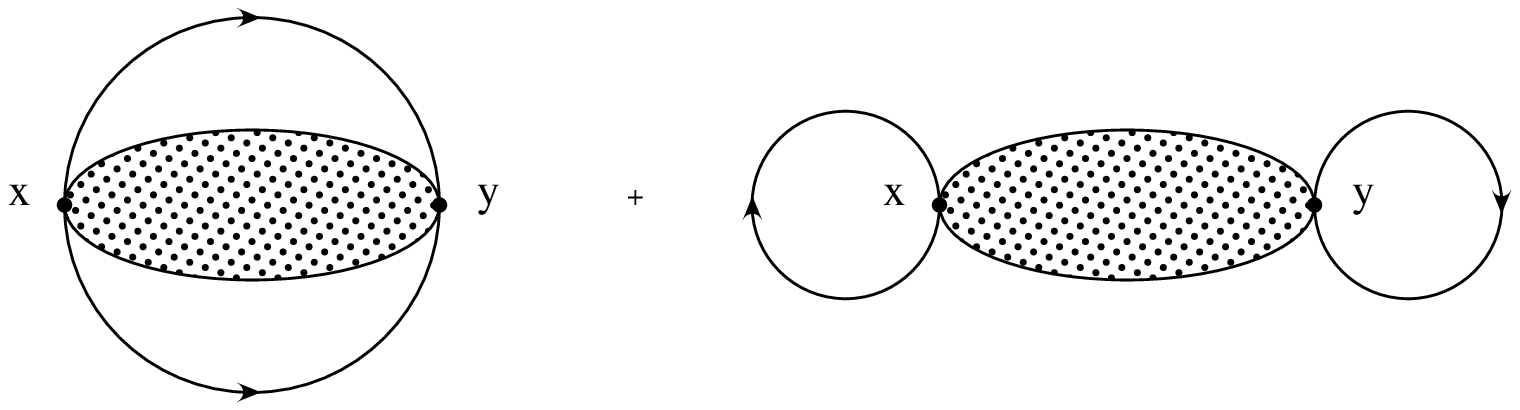, width=\linewidth}
\end{center}
\caption{Vacuum bubbles}
\label{fig2}
\end{figure}
\begin{figure}
\begin{center}
\epsfig{file=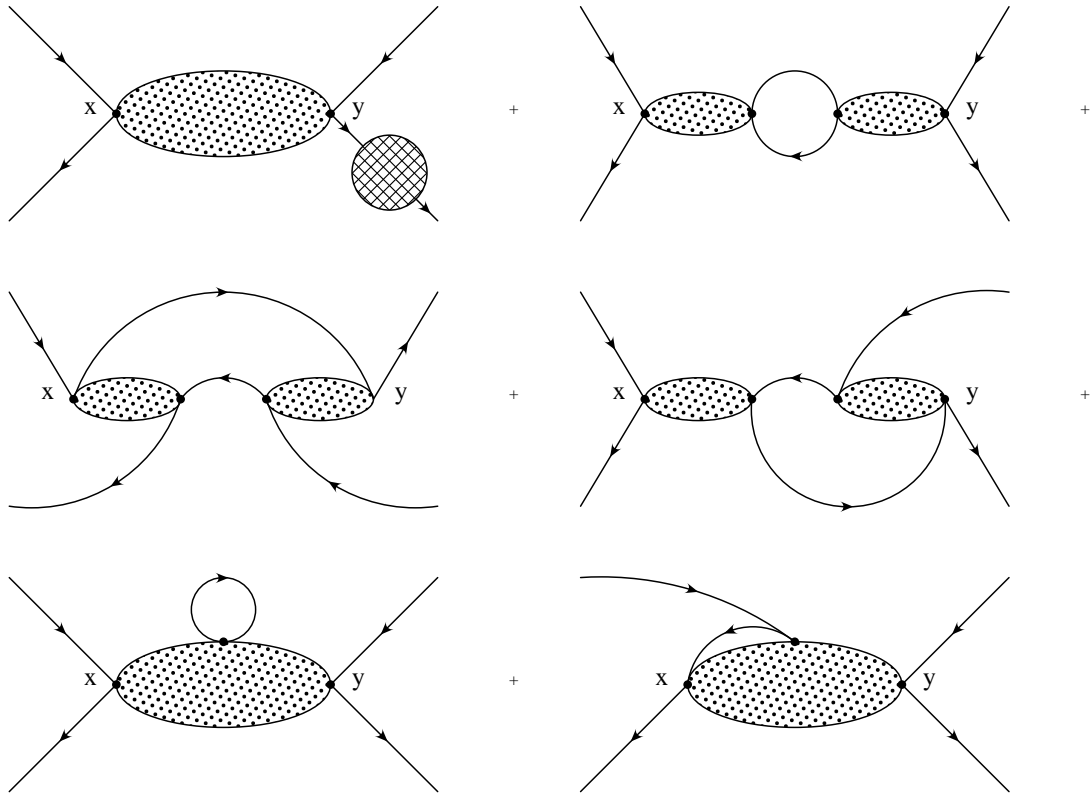, width=\linewidth}
\end{center}
\caption{Lowest order vertex corrections}
\label{fig4}
\end{figure}

Other loop contributions are vacuum bubbles (shown in fig.\ \ref{fig2}) and vertex corrections (shown in fig.\ \ref{fig4}). 
Another remark concerns the non-local loops (the basketball-like graph in fig.\ \ref{fig2} or the sunset-like graph in fig.\ \ref{fig3}). Their appearance is a consequence of the effective non-local interactions present in (\ref{Wtilde-b}) (when (\ref{pr:20}) is inserted). An important question is whether they are relevant or not. We conjecture that they disappear in general, although we have checked this statement only for particular cases and only to lowest order. This conjecture is based upon the following observations:
\blist
\item The classical 4-point vertices vanish in the coincidence limit $x^0\to y^0$ for all GDTs (cf.\ eqs.\ (3.26) and (3.27) of \cite{Grumiller:2002dm})
\item The vertices are local in $x^1$, i.e.\ they are proportional to $\de(x^1-y^1)$ (cf.\ eqs.\ (3.26) and (3.27) of \cite{Grumiller:2002dm})
\item The internal propagator yields a $\de$-function (or some derivative thereof) in $x^1+g(x^0)-y^1-g(y^0)$, where $g$ is a known function for each model (cf.\ eqs.\ (\ref{1loop3}) and (\ref{1loop4}) below)
\elist
It would be interesting to either prove this conjecture (maybe from some general principles) or to find a counter example, because it would rule out many graphs which are otherwise difficult to calculate (e.g.\ the third, fourth and sixth graph of fig.\ \ref{fig4}).

In the following we would like to shed some light on the third statement in the list above for it is the least trivial one.

Performing the Gaussian integral over $\de\phi$ with
the measure $(df^{1/2}\de\phi )$ of the exponentiated
first line of (\ref{Lsig2}) one obtains the Polyakov
action ${\cal L}_P(f,\Emo )$, where, as seen
from (\ref{Lsig2})
\begin{equation}
\Emo =\nabla_0^{-2} \left[ -w \right]''(\hat{X}(\phi_0))=(M+w(\hat{X}(\phi_0)))\,.
\label{moreEmo}
\end{equation}
According to (\ref{cghs:4}), the conformal factor $f$
has to be replaced by its value of order $\hbar^0$:
\begin{equation}
f=\Epo = I(\hat{X}(\phi_0)) \,. \label{moreEpo}
\end{equation}

This representation of the one-loop diagrams through the
Polyakov action on the background of some effective geometry 
has two obvious advantages. First, a closed
renormalized expression for that loop is available. Second,
this representation facilitates comparison with perturbative
calculations of $2D$ Hawking radiation which are usually done
with the help of the Polyakov action.

The Polyakov action in Eddington-Finkelstein gauge is
\begin{eqnarray}
&&{\cal L}_P(E_1^+,E_1^-)=-\frac{1}{96\pi}\int_x \int_y
(\partial_0^2 E_1^- -\Gamma \ln E_1^+)_x\Gamma^{-1}_{xy} 
(\partial_0^2 E_1^- -\Gamma \ln E_1^+)_y \nonumber \\
&&\qquad \Gamma = \partial_1\partial_0 -\partial_0 E_1^-
\partial_0 \label{Lpol-gauge}
\end{eqnarray}
Functional derivatives of (\ref{Lpol-gauge}) with
respect to $E^\pm_1$ describe closed matter loops
with an external line. After calculating the derivatives
the fields $E^\pm_1$ must be taken at the 
(e.g.\ Minkowski) background. The equation
\begin{equation}
\frac{\delta}{\delta E_1^\pm} {\cal L}_P|_{\rm BG}=0
\label{stabM}
\end{equation}
expresses stability of the background space and
guarantees the absence of tadpoles.
The ``$+$'' component of the equation (\ref{stabM})
is reduced to the (trivial) equation:
\begin{equation}
\Gamma \ln E_1^+=\partial_0^2 E_1^- \ \label{GamEE}
\end{equation}
The ``$-$'' component of (\ref{stabM}) reads
\begin{eqnarray}
&& 2\partial_0 {\cal E}(x) -{\cal E}(x)^2 = 2\partial_0^2\ln{E_1^+}-(\partial_0\ln{E_1^+})^2 \nonumber \\ 
&& {\cal E}(x):=\partial_0^x \int_y \Gamma^{-1}_{xy} \partial_{0}^2 E_1^-(y)
\label{123}
\end{eqnarray}
This equation is satisfied if
\begin{equation}
\partial_0 \ln E_1^+=
\partial_0 \Gamma^{-1}\partial_0^2 E_1^- \ \label{EGamE}
\end{equation}  
The equation (\ref{EGamE}) defines the action of
$\partial_0 \Gamma^{-1}$ on functions of $x^0$.
The operator $\partial_0 \Gamma^{-1}$ can be represented as
\begin{equation}
(\partial_0 \Gamma^{-1})_{xy}=\frac{1}{E_1^-(x^0)} 
(\gamma^{-1})_{xy} E_1^-(y^0) \,, \label{dGam}
\end{equation}
where
\begin{equation}
\gamma =\partial_1 - E_1^-\partial_0\,.
\label{gamma}
\end{equation}
It is convenient to introduce holonomic light-cone coordinates $u,v$, defined by
\eq{
u \propto x^1\,,\quad v \propto x^1+\int_{x^0} \frac{dz}{E_1^-(z)}\,,
}{1loop3}
since (\ref{gamma}) then simplifies to $\ga\propto\partial_u$ (note incidentally that classically $E_1^-=-M-w(x^0)$ holds -- i.e.\ the usual Regge-Wheeler coordinate transformation is employed). Then eq.\ (\ref{gamma}) implies the localization of $\ga^{-1}_{xy}$ in $v$, i.e.\
\eq{
(\partial_0 \Gamma^{-1})_{xy}\propto\de(v-v^0)\,.
}{1loop4}
This concludes the proof of the third statement above, at least to leading order in a loop expansion, for all GDTs\footnote{An important caveat concerns the appearance of distributions and derivatives thereof with $u-u_0$ as argument in (\ref{1loop4}). They do not spoil the argument at one-loop level, since the zeros of the vertices are ``strong'' enough to compensate, but it is conceivable that at higher loop order or for higher order $2n$-point functions this line of reasoning breaks down.}.

If our conjecture is true indeed, only local loop interactions contribute and we have the usual QFT situation, except that our tree-level vertices are non-local.

\section{Twenty questions and answers}\label{app:4}

These questions occurred either during the lectures or afterwards in private discussions. I am grateful to all students and colleagues who asked (some of) them.
\begin{enumerate}
\item We have seen that the vacuum expectation values for the Cartan variables coincide with their classical values ($<e_\mu^\pm>=e_\mu^\pm$, $<\om>=\om$); however we have not addressed correlators yet; the most relevant seems to be the metric; {\em Q:} is $<e_\mu^+e_\nu^->=<e_\mu^+><e_\nu^->=e_\mu^+e_\nu^-$ or not? {\em A:} it is not the same; thus, although the vielbeine behave classically the metric does not; this seems to be in contradiction with the formal off-shell equivalence of first and second order formalism of gravity \cite{Aros:2003bi}, but that proof is restricted to Einstein-Hilbert in $4D$, while we are dealing with dilaton gravity in $2D$; so to summarize, it seems that for quantization the Cartan formulation is technically superior as compared to the metric formulation.
\item {\em Q:} What about expectation values of geodesics? {\em A:} in geodesics the Killing norm $\xi$ appears (which is quadratic in the vielbeine); thus, the previous issue is also relevant here and we get something like ``quantum corrections to geodesics''; however, these corrections are singular, so this just seems to indicate that one should test geometry with fields rather than with point particles at the quantum level.
\item {\em Q:} What about positivity of the dilaton in the path integral? {\em A:} we have integrated over all paths, i.e.\ also over negative dilatons; a ``physically motivated'' path integration would have restricted the dilaton to the positive half-line -- but then we would have obtained instead of a $\de$-function only half the $\de$-function and a principal value contribution (similar to QED when you take the ``physically motivated'' retarded prescription instead of the causal one); only the $\de$-function yields the correct classical limit.
\item {\em Q:} What is the dimensional dependence of the S-matrix for $D$-dimensional s-wave gravitational scattering? {\em A:} I don't know how to answer this without performing too many calculations, but maybe one can use this trick: assuming finiteness of the S-matrix for all dimensions it seems sufficient to calculate one of the two vertices and truncate the divergent part; for simplicity, one can take the non-symmetric vertex, since it involves only $F\propto X$, $(x^0-y^0)$, $\phi_0$ and $\phi_1$, i.e.\ the potentials are felt only indirectly via the asymptotic states.
\item {\em Q:} What does the vanishing of the VBH (or the vertices) in the coincidence limit $x^0\to y^0$ mean? {\em A:} this has important implications for non-local loops because it makes them vanish; see the discussion below (\ref{Lsig2}).
\item {\em Q:} Does the VBH exist for all GDTs? {\em A:} yes, almost (see \cite{Grumiller:2002dm}). 
\item {\em Q:} What about Hawking radiation of VBHs? {\em A:} this issue is problematic to address because although the Ricci scalar of the VBH is a regular distribution the flux component is singular (contains typically terms of the form $\xi\xi''-(\xi')^2$); so it is probably pointless to attribute a (Hawking) temperature to a VBH.
\item {\em Q:} What about the notion of ``light-like'' -- does it become ``fuzzy'' or is it still valid to talk about ``light-like'' separations in the quantum case? {\em A:} within our approach I see no way how the light-cone could become fuzzy.
\item {\em Q:} What about non-smooth potentials? {\em A:} they lead to non-smooth geometries; depending on the degree of non-smoothness one can use a twodimensional analogon of the Israel junction conditions.
\item {\em Q:} What about the relation to noncommutative geometry? {\em A:} in noncommutative geometry: $[X^i,X^j]=\theta^{ij}(X)$; we have this structure on the Poisson manifold (i.e.\ our target space with coordinates $X^i(x)$), but nothing analogous on the worldsheet.
\item {\em Q:} What about the Weyl curvature hypothesis in the context of VBHs? {\em A:} first of all, this question makes only sense for dimensionally reduced gravity, because there is no intrinsically twodimensional Weyl tensor; but even referring to the higherdimensional quantities the problem is that $C^{abcd}C_{abcd}$ is ill-defined for a VBH geometry; on the other hand, $R^{abcd}R_{abcd}$ can be calculated and it vanishes, if integrated -- so if we would take this at face value this would imply a ``high'' (namely divergent) gravitational entropy.
\item {\em Q:} What about non-perturbative issues? {\em A:} one can reconstruct geometry out of any given matter solution; we have performed this perturbatively up to now; if one can invent a full nonperturbative solution for the scalar field then one can easily perform the same reconstruction; the problem is, of course, to find such a solution. Classically, but not quantum mechanically, the CGHS \cite{Callan:1992rs} is soluble exactly (and with certain modifications also semiclassically, cf.\ \cite{Zaslavsky:1998ca} and references therein).
\item {\em Q:} What about the signature of the metric in the path integral formalism -- can it change? {\em A:} in our approach signature is fixed automatically and cannot change (apart from singular points where the metric degenerates).
\item There seems to be a double counting of paths: if we let $e^\pm_\mu\to-e^\pm_\mu$ and $\om\to-\om$ the whole action changes sign; the equations of motion are invariant under this discrete trafo; thus, we have a (classical) $\mathbb{Z}_2$ symmetry --{\em Q:} don't we count everything twice? {\em A:} no, because similar to point 3 we must sum over all of these paths or else we don't get $\de$-functions! note, however, that the boundary conditions we employ break the $\mathbb{Z}_2$ symmetry (any non-singular ($\sqrt{-g}\neq 0$) vacuum will break this symmetry); this ``explains'' why even the trivial boundary conditions we have used ($M=0$) can yield a non-trivial S-matrix.
\item {\em Q:} Why is there no ordering problem? {\em A:} essentially because of the linearity in the 1-components of the Cartan variables in the action. 
\item {\em Q:} Is there a point mechanical model resembling dilaton gravity? {\em A:} unfortunately I don't know of any -- one would need a Hamiltonian of the form $H(p^j,q_j)=q_if^i(p_j)$, where $f^i$ are arbitrary functions of the ``momenta''; such a Hamiltonian has some similarity to (part of the) Runge-Lentz term, but models with this term always contain some $p^2$ terms or $q^2$ terms.
\item {\em Q:} Aren't the VBHs examples of ``nullitons'' (a lightlike analogon of solitons)? {\em A:} yes, but the VBH corresponds to a singular field configuration, so no (cf.\ the discussion in \cite{Fischer:2001vz}); one could turn the question around and ask ``What is a lightray of finite (affine) length''? if one imposes regularity conditions there doesn't seem to exist a meaningful solution \cite{Visser:2003na}.
\item According to standard folklore quantum gravity should allow for topology changes {\em Q:} does it really happen? {\em A:} the simplest of such fluctuations would be a change of the Euler-characteristic; look at the VBH: if we cut out everything that is singular (i.e. the VBH) we have Minkowski-space time with something cut out; in the Carter-Penrose-diagram we have cut out 2 points and 1 line, so the Euler-number changes by 1 unit; for spherically reduced gravity, on each point (expect for $r=0$) there sits an $S^2$, but again one can show that the Euler number changes by 1 unit; so VBHs, viewed as intermediate states, really seem to fit into the picture of ``fluctuating topology''; on the other hand, we have emphasized that the VBH geometry (which is an off-shell quantity) should not be taken at face value; in the ``real'' geometry no topology changes do occur because we have imposed certain boundary conditions in order to obtain our S-matrix results -- so our asymptotic observer does not observe any topology change.
\item {\em Q:} In what sense the effective theory one obtains after integrating out geometry is nonlocal? {\em A:} The formal answer would be: there appear terms of the form $\nabla_0^{-2}$, i.e.\ nonlocal operators, in the effective action. A more physical answer can be given by examination of the essential part of the S-matrix (\ref{feynman}): it is polynomial in the energy (in fact, it is even independent of the total energy), but obviously nonpolynomial in the momenta. One of the standard definitions of locality (cf. e.g.\ \cite{peskinquantum}) is polynomiality in the momenta. In this sense, our S-matrix is nonlocal.
\item {\em Q:} What about a boundary term a la Kucha{\v{r}} \cite{Kuchar:1994zk} -- does it appear somehow in the path integral? {\em A:} Not straightforwardly. There is indeed a boundary term in the effective action (cf.\ eq.\ (7.55) of \cite{Grumiller:2002nm}). And the mass $M$ is hidden in that term which in the present notation reads $\mp\int_{\partial M}dx^1g(x^1)(M+w(x^0_B))$ -- however, there is an additional term which typically diverges at $x^0_B=\infty$; one could impose a subtraction (i.e.\ to subtract the value of the boundary term for $M=0$) and thus obtains a renormalized boundary term which essentially coincides with Kucha{\v{r}}'s result $\int_{\partial M}dt \dot\tau M$, provided one identifies $dx^1g(x^1)=dt\dot\tau$. However, this seems to be a rather indirect way as compared to Kucha{\v{r}}'s derivation -- so although the path integral seems to be the most adequate language to describe scattering processes other aspects might be obscured a bit.
\end{enumerate}

\end{appendix}

%%% END OF PAPER %%%

%%% REFERENCES %%%

%\bibliographystyle{../../projekte/review01/fullsort} % for the author's convenience
%\bibliography{../../projekte/review01/review} % my bibtex-file - just use the .bbl file instead or I can send you my bibtex file, if you want (but please do not include new refs in it - rather let me include them, it is much easier to handle this ``centralized''

\providecommand{\href}[2]{#2}\begingroup\raggedright\endgroup

%%% END OF REFERENCES %%%

\end{document}